%
%
%
%
\input harvmac

\def\C{{\bf C}}

\def\a{\alpha}

\def\l{\lambda}

\def\m{\mu}

\def\f{\phi}
\def\F{\Phi}
\def\w{\omega}

\def\P{{\bf P}}

\def\d{\partial}

\def\inv{^{-1}}
\def\Tr{{\rm Tr}}

\def\cO{{\cal O}}
\def\cF{{\cal F}}
\def\cN{{\cal N}}

\lref\thooft{
G.~'t Hooft, ``A Planar Diagram Theory For Strong Interactions,''
Nucl.\ Phys.\ B {\bf 72}, 461 (1974).}

\lref\mm{
P.~Ginsparg and G.~W.~Moore,
``Lectures On 2-D Gravity And 2-D String Theory,''
arXiv:hep-th/9304011.}

\lref\mmm{
P.~Di Francesco, P.~Ginsparg and J.~Zinn-Justin,
``2-D Gravity and random matrices,''
Phys.\ Rept.\  {\bf 254}, 1 (1995)
[arXiv:hep-th/9306153]}

\lref\kontsevich{M.~Kontsevich,
``Intersection Theory On The Moduli Space Of Curves And The Matrix
Airy Function,'' Commun.\ Math.\ Phys.\ {\bf 147}, 1 (1992).}

\lref\wittentop{E.~Witten,
``On The Structure Of The Topological Phase Of Two-Dimensional
Gravity,'' Nucl.\ Phys.\ B {\bf 340}, 281 (1990)}

\lref\matrix{T.~Banks, W.~Fischler, S.~H.~Shenker and L.~Susskind,
``M theory as a matrix model: A conjecture,''
Phys.\ Rev.\ D {\bf 55}, 5112 (1997)
[arXiv:hep-th/9610043].}

\lref\adscft{
O.~Aharony, S.~S.~Gubser, J.~M.~Maldacena, H.~Ooguri and Y.~Oz,
``Large N field theories, string theory and gravity,''
Phys.\ Rept.\  {\bf 323}, 183 (2000)
[arXiv:hep-th/9905111].
}

\lref\ghv{D. Ghoshal and C. Vafa, ``c=1 string as the topological theory
of the conifold,''
Nucl.\ Phys.\ B {\bf 453}, 121 (1995)
[arXiv:hep-th/9506122].}

\lref\gv{R.~Gopakumar and C.~Vafa,
``On the gauge theory/geometry correspondence,''
Adv.\ Theor.\ Math.\ Phys.\  {\bf 3}, 1415 (1999)
[arXiv:hep-th/9811131].
}

\lref\edel{J.D. Edelstein, K. Oh and R. Tatar, ``Orientifold,
geometric transition and large $N$ duality for SO/Sp gauge theories,''
JHEP {\bf 0105}, 009 (2001)
[arXiv:hep-th/0104037].}

\lref\dasg{K. Dasgupta, K. Oh and R. Tatar, ``Geometric transition, large
$N$ dualities and MQCD dynamics,''  Nucl. Phys.
B {\bf 610}, 331 (2001)
[arXiv:hep-th/0105066]\semi ``Open/closed string dualities and Seiberg
duality from geometric transitions in M-theory,''
[arXiv:hep-th/0106040]\semi ``Geometric transition versus cascading
solution,''
JHEP {\bf 0201},  031 (2002)
[arXiv:hep-th/0110050].}

\lref\vaug{C.~Vafa,
``Superstrings and topological strings at large N,''
J.\ Math.\ Phys.\  {\bf 42}, 2798 (2001)
[arXiv:hep-th/0008142].}

\lref\civ{
F.~Cachazo, K.~A.~Intriligator and C.~Vafa,
``A large N duality via a geometric transition,''
Nucl.\ Phys.\ B {\bf 603}, 3 (2001)
[arXiv:hep-th/0103067].}

\lref\ckv{
F.~Cachazo, K.~A.~Intriligator and C.~Vafa, ``A large N duality via a
geometric transition,'' Nucl.\ Phys.\ B {\bf 603}, 3 (2001)
[arXiv:hep-th/0103067].  }

\lref\cfikv{
F.~Cachazo, B.~Fiol, K.~A.~Intriligator, S.~Katz and C.~Vafa,
``A geometric unification of dualities,'' Nucl.\ Phys.\ B {\bf 628}, 3
(2002) [arXiv:hep-th/0110028].}

\lref\cv{
F.~Cachazo and C.~Vafa, ``N = 1 and N = 2 geometry from fluxes,''
arXiv:hep-th/0206017.}

\lref\ov{
H.~Ooguri and C.~Vafa, ``Worldsheet derivation of a large N duality,''
arXiv:hep-th/0205297.}

\lref\av{
M.~Aganagic and C.~Vafa, ``$G_2$ manifolds, mirror symmetry,
and geometric engineering,'' arXiv:hep-th/0110171.}

\lref\digra{
D.~E.~Diaconescu, B.~Florea and A.~Grassi, ``Geometric transitions and
open string instantons,'' arXiv:hep-th/0205234.}

\lref\amv{
M.~Aganagic, M.~Marino and C.~Vafa,
``All loop topological string amplitudes from Chern-Simons theory,''
arXiv:hep-th/0206164.}

\lref\dfg{
D.~E.~Diaconescu, B.~Florea and A.~Grassi, ``Geometric transitions,
del Pezzo surfaces and open string instantons,''
arXiv:hep-th/0206163.}

\lref\kkl{
S.~Kachru, S.~Katz, A.~E.~Lawrence and J.~McGreevy,
``Open string instantons and superpotentials,''
Phys.\ Rev.\ D {\bf 62}, 026001 (2000)
[arXiv:hep-th/9912151].}

\lref\bcov{
M.~Bershadsky, S.~Cecotti, H.~Ooguri and C.~Vafa, ``Kodaira-Spencer
theory of gravity and exact results for quantum string amplitudes,''
Commun.\ Math.\ Phys.\ {\bf 165}, 311 (1994) [arXiv:hep-th/9309140].
}

\lref\witcs{
E.~Witten,
``Chern-Simons gauge theory as a string theory,''
arXiv:hep-th/9207094.
}

\lref\shenker{
S.~H.~Shenker, ``The Strength Of Nonperturbative Effects In String
Theory,'' in Proceedings Cargese 1990, {\it Random surfaces and
quantum gravity}, 191--200.  }

\lref\berwa{
M.~Bershadsky, W.~Lerche, D.~Nemeschansky and
N.~P.~Warner, ``Extended N=2 superconformal structure of gravity and W
gravity coupled to matter,'' Nucl.\ Phys.\ B {\bf 401}, 304 (1993)
[arXiv:hep-th/9211040]}

\lref\loop{
G.~Akemann, ``Higher genus correlators for the Hermitian matrix model
with multiple cuts,'' Nucl.\ Phys.\ B {\bf 482}, 403 (1996)
[arXiv:hep-th/9606004].  }

\lref\wiegmann{P.B. Wiegmann and A. Zabrodin, ``Conformal maps
and integrable hierarchies,'' arXiv:hep-th/9909147.}

\lref\cone{R.~Dijkgraaf and C.~Vafa, to appear.}

\lref\kazakov{
S.~Y.~Alexandrov, V.~A.~Kazakov and I.~K.~Kostov,
``Time-dependent backgrounds of 2D string theory,''
arXiv:hep-th/0205079.}

\lref\dj{
S.~R.~Das and A.~Jevicki,
``String Field Theory And Physical Interpretation Of D = 1 Strings,''
Mod.\ Phys.\ Lett.\ A {\bf 5}, 1639 (1990).}

\lref\givental{
A.B.~Givental, ``Gromov-Witten invariants and quantization of quadratic hamiltonians,''
 arXiv:math.AG/0108100.}

\lref\op{
A.~Okounkov and R.~Pandharipande, ``Gromov-Witten theory, Hurwitz
theory, and completed cycle,'' arXiv:math.AG/0204305.}

\lref\dijk{ R.~Dijkgraaf, ``Intersection theory, integrable hierarchies and
topological field theory,'' in Cargese Summer School on {\it New Symmetry
Principles in Quantum Field Theory} 1991,
[arXiv:hep-th/9201003].}

\Title
 {\vbox{
 \hbox{hep-th/0206255}
 \hbox{HUTP-02/A028}
 \hbox{ITFA-2002-22}
}}
{\vbox{
\centerline{Matrix Models, Topological Strings, and}
\vskip 5mm
\centerline {Supersymmetric Gauge Theories}
}}
\centerline{Robbert Dijkgraaf}
\vskip.05in
\centerline{\sl Institute for Theoretical Physics \&}
\centerline{\sl Korteweg-de Vries Institute for Mathematics}
\centerline{\sl University of Amsterdam}
\centerline{\sl 1018 TV Amsterdam, The Netherlands }
\smallskip
\centerline{and}
\smallskip
\centerline{Cumrun Vafa}
\vskip.05in
\centerline{\sl Jefferson Physical Laboratory}
\centerline{\sl Harvard University}
\centerline{\sl Cambridge, MA 02138, USA}

\vskip .1in\centerline{\bf Abstract}

\smallskip

We show that B-model topological strings on local Calabi-Yau
threefolds are large $N$ duals of matrix models, which in the planar
limit naturally give rise to special geometry. These matrix models
directly compute F-terms in an associated ${\cal N}=1$ supersymmetric
gauge theory, obtained by deforming ${\cal N}=2$ theories by a
superpotential term that can be directly identified with the potential
of the matrix model.  Moreover by tuning some of the parameters of the
geometry in a double scaling limit we recover $(p,q)$ conformal
minimal models coupled to 2d gravity, thereby relating non-critical
string theories to type II superstrings on Calabi-Yau backgrounds.

\Date{June, 2002}


\newsec{Introduction}

Large $N$ limits of $U(N)$ gauge theories have been a source of
inspiration in physics, ever since 't Hooft introduced the idea
\thooft.  In particular the large $N$ limit of gauge theories should
be equivalent to some kind of closed string theory.  The first contact
this idea had with string theory was in the context of non-critical
bosonic strings described by $c \leq 1$ conformal field theories
coupled to two-dimensional gravity.  It was found that by taking a
``double scaling limit'' of $N\times N$ matrix models, where one send
$N$ to infinity while at the same time going to some critical point,
one ends up with non-critical bosonic strings
\refs{\mm,\mmm}. This relation between gauge systems and strings was not
exactly in the sense that 't Hooft originally suggested for the large
$N$ expansion, as it involved a double scaling limit.  In particular,
before taking this limit the matrix model would not have a string
dual, whereas according to 't Hooft's general idea one would have
expected it to have.

In one context this was remedied by a different kind of matrix model
introduced by Kontsevich \kontsevich, where without taking a double
scaling limit one finds an equivalence between a matrix model and
non-critical string theory. In particular the amplitudes of the
topological string observables introduced in \wittentop\ are directly
computed by these matrix integrals.  This duality was in the same
spirit as 't Hooft's general idea and can be seen as a low-dimensional
example of a holographic correspondence.

More recently large $N$ dualities have come back in various forms.
In the context of M-theory a large $N$ matrix formulation was advanced
\matrix.  Since here an unconventional large $N$ limit is involved,
this again was not quite in the same spirit as 't Hooft's idea, much
as the double scaling limit of matrix models in the context of
non-critical strings is not.  However, the AdS/CFT correspondence
\adscft\ is in the same spirit as 't Hooft's original
proposal in that one did not have to take a particular limit to obtain
an equivalence.  Another example of such a strict large $N$ duality is
the relation between Chern-Simons gauge theory and A-model topological
strings \gv\ where one also does not have to take any particular limit
for the equivalence to hold.

The main aim of this paper is to develop a mirror version of this last
duality \gv.  We find matrix models that are dual to B-model
topological strings on Calabi-Yau threefolds.  This is again in the
same spirit as 't Hooft, as one does not take a double scaling
limit. We will show in particular that the special geometry of
Calabi-Yau threefolds that solves the B-model at tree level emerges
naturally from the dynamics of the eigenvalues of the matrix
model. 

However, even though it is not required, one {\it can} also consider a
double scaling limit of this setup and obtain a specific class of
Calabi-Yau manifolds that are dual to double-scaled matrix models.  In
this sense we are enlarging the original equivalence of double scaling
limits of matrix models with string theory to an equivalence of all
matrix models with some kind of closed string theory, without any need
to take a double scaling limit. In particular our result shows that
studying strings on non-compact Calabi-Yau spaces provides a unifying
approach to all matrix model descriptions.

Furthermore, it turns out that one can embed these large $N$ dualities
in the context of type IIB superstrings \vaug, a relation that was
further explored in \refs{\civ, \edel, \dasg, \ckv,\cfikv,\cv}.  In
the context of this embedding one obtains a dictionary in which the
planar limit of the matrix models is seen to compute superpotentials
for certain ${\cal N}=1$ supersymmetric gauge theories, where the
potential of the matrix model gets mapped to the superpotential for an
adjoint scalar of an ${\cal N}=1$ theory.

The organization of this paper is as follows: In section 2 we propose
the large $N$ conjecture for topological strings with matrix models
after reviewing the various geometrical ingredients.  In section 3 we
pass this conjecture through some highly non-trivial checks, and in
particular check it at the planar limit.  In section 4 we discuss some
generalizations of this conjecture and its connections with
non-critical bosonic strings coupled to gravity and the double scaling
limit.  We also discuss the meaning of the double scaling limit from
the viewpoint of type IIB superstrings.

\newsec{Large $N$ Topological String Conjectures}

In this section we formulate a general class of large $N$ conjectures
involving B-model topological strings and certain two-dimensional
topological gauge theories.  The idea is to consider the mirror of the
large $N$ conjecture of \gv, which relates large $N$ Chern-Simons
theory on $S^3$ with A-model topological strings on the resolved
conifold.  This duality involves Calabi-Yau threefold transitions
where a 3-cycle with branes wrapped over it shrinks and a 2-cycle
grows without any branes wrapped over it.  A worldsheet derivation of
this duality has been recently presented in \ov.  Moreover this has
been generalized to a large class of Calabi-Yau threefolds \av\ which
has been further studied in \digra\ leading to development of powerful
methods to compute all loop A-model topological string amplitudes
\refs{\amv,\dfg}.

As suggested in \gv\ the mirror of A-model transitions dualities
should also exist, namely we can consider 2-cycles with branes wrapped
over them, undergoing transitions where they shrink and are replaced
by 3-cycles without any branes.  A large class of such Calabi-Yau
transitions were considered in \refs{\civ,\ckv,
\edel,\dasg,\cfikv}\ in the context of
embedding such B-model dualities in type IIB superstrings.  However,
the duality between gauge theory and the topological B-model itself,
which is implicit in these works, has not been studied.  The aim of
this section is to elaborate on these topological B-model/large $N$
gauge theory dualities.

\subsec{Geometry of the generalized conifold transition}

Instead of being general we consider a special class of such
transitions, studied in \civ , and discuss its topological lift.  The
situation considered in \civ\ involved a string theory realization of
${\cal N}=2$ supersymmetric $U(N)$ gauge theory, deformed to ${\cal
N}=1$ theory by addition of a tree-level superpotential ${\rm Tr}\,
W(\Phi )$, which we take to be a general polynomial of degree $n+1$
of the adjoint field $\Phi$.
The Calabi-Yau geometry relevant for this was studied in
\kkl\ and corresponds to considering
the blowup of the local threefold given as a hypersurface
\eqn\sinc{uv+y^2+W'(x)^2=0.}
The blow up takes place at the critical points of $W$, {\it i.e.} at
$W'(x)=0$.  Such transitions lead to a geometry that contains $n$
blown up $\P^1$'s which are all in the same homology class. So in the
resolved geometry we can find $n$ isolated rational curves.

More precisely, the resolved singularity can be obtained by starting
with the bundle $\cO(0) \oplus \cO(-2)$ over $\P^1$.  This is the
normal bundle to rational curve in $K3 \times \C$ and corresponds to
${\cal N}=2$ supersymmetry. Let us denote the sections of the normal
bundle by $\f_0$, $\f_1$. These sections are respectively 0-forms and
1-forms on the $\P^1$. The field $\f_0$ corresponds to the adjoined
valued Higgs field $\F$ in the $\cN=2$ SYM theory.

Now the inclusion of the superpotential $W$ should give $n$ isolated
$\P^1$'s at the critical values $W'(\f_0)=0$. This is achieved by the
following transition function. If $z$ and $z'$ are the coordinates on
the northern and southern hemispheres of the $\P^1$, then the
resolution is given by relating the patches $(z,\f_0,\f_1)$ and
$(z',\f_0',\f_1')$ by
\eqn\patch{
\eqalign{
\f_0' & = \f_0, \cr
z' \f_1' & = z \f_1 + W'(\f_0), \cr
z' & = 1/z. \cr
}}
To relate it to the geometry given by \sinc\ we make the identifications
\eqn\relc{
\eqalign{
x & =\f_0, \cr
u & =2\f_1',\cr
v & =2\f_1,\cr
\omega & =z'\f_1',\cr
y & = i(2\omega -W'(\f_0)), \cr
}}
Here we have introduced another variable $\omega$ in terms of which
the geometry would have been given by
\eqn\omeg{
uv-4\omega^2+4\omega W'(x)=0.
}
 This makes a natural connection with objects that will be introduced
in the next section.

If we now distribute $N$ D5 branes
wrapped over these $S^2$'s and filling the spacetime, this corresponds
to a choice of the vacuum in the corresponding ${\cal N}=1$
supersymmetric gauge theory, where the distribution of the branes
among the $n$ critical points corresponds to distributing the $N$
eigenvalues of $\Phi$ among the $n$ classical values.  At large $N$
gaugino condensation takes place and this leads to a geometric
transition in which the $S^2$'s are all blown down and replaced by
``blown up'' $S^3$'s.  This results in the geometry \civ\
\eqn\lar{uv+y^2+W'^2(x)+f(x)=0,}
where $f(x)$ is a polynomial of degree $n-1$, whose precise
coefficients depend on the distribution of the $N$ eigenvalues among
the $n$ critical points.  

In \cv\ it has been shown that, if one choses $W(x)$ to be of degree
$N+1$ and if all the critical points are equally occupied, the ${\cal
N}=2$ Seiberg-Witten geometry can be recovered by considering the
limit $W\rightarrow \epsilon W$ as $\epsilon \rightarrow 0$.

Note in particular the identification of
$x$ with the eigenvalues of $\Phi$. This is rather natural from
the equation \sinc .  In particular if $W=0$ then we have an $A_1$
geometry formed as a hypersurface in $(u,v,y)$ space, where we have
wrapped a $D5$ brane around the blowup sphere.  The transverse
position of the $D5$ brane along the $x$ direction would correspond to
changing the vev for $\Phi$.  Having a non-trivial fibration of the
$A_1$ geometry over the $x$-plane dictated by \sinc\ gives rise to the
superpotential $W(\Phi)$, as reviewed above.

We will be interested in the B-model topological string on the
deformed geometry given by equation \lar. The genus zero prepotential
is determined by period integrals of the holomorphic (3,0) form
$$
\Omega = {dx\wedge du\wedge dv \over \sqrt{uv + W'(x)^2 + f(x)}}.
$$
This non-compact Calabi Yau has $n$ compact three-cycles $A_i$ that
correspond to the $n$ ``blown-up'' $S^3$ and there are $n$ canonically
conjugated cycles $B_i$ that are non-compact and have the topology of
a three-ball. The usual special geometry relations determine the
tree-level prepotential $\cF_0(S_i)$ by the periods
\eqn\periods{
\eqalign{
S_i & = \int_{A_i} \Omega, \cr
{\d \cF_0 \over \d S_i} & = \int_{B_i} \Omega. \cr
}}
Here one should be careful in regulating the periods over the
non-compact $B$ cycles \civ. The degree of the deformation $f(x)$, which
corresponds to normalizable deformations, is such
that these periods can be sensibly defined.  In particular their
variations  with respect to the variation of the coefficients of $f(x)$
are cutoff independent.

As explained in \civ\ the $A$ and $B$ cycles can be represented as
$S^2$ fibrations over paths in the complex $x$-plane. After
integrating the 3-form over these $S^2$ fibers we are left with
the integrals over these curves of the 1-form
$$
\eta = y dx.
$$
Here $y$ is determined by the hyperelliptic curve
\eqn\cycurve{
y^2 + W'(x)^2 + f(x) = 0.
}
We will refer to this curve as the
spectral curve.  It has $n$ branch cuts that are the projections of
the cycles $A_i$ onto the $x$-plane. The cycles $B_i$ are represented
by half-lines that start at the cuts and run to some cut-off point
$x=\Lambda$ far away from the branch cuts.

\subsec{Lifting to topological string dualities}

We now consider lifting these dualities to topological strings.  As
discussed in \vaug\ the key point is the observation in
\bcov\ that topological strings computes superpotential terms
of the corresponding gauge theory arising in type II superstrings.  In
particular the leading planar diagram computes superpotential terms
involving the gaugino bilinear field on the gauge theory side, and the
${\cal N}=2$ prepotential on the dual gravity side (with some vev's
for auxiliary fields, corresponding to turning on fluxes).  More
generally the topological gravity in the presence of Calabi-Yau 3-fold
is the B-model theory studied in \bcov\ and called ``Kodaira-Spencer
theory of gravity''.  Thus all we need to do is to specify the gauge
theory dual, which should be the gauge theory on the topological
branes wrapping the ${\bf P}^1$'s.  This theory is the reduction of
the holomorphic Chern-Simons theory studied in \witcs\
from complex dimension three to complex
dimension one, which we now turn to.

First suppose that we have no superpotential, {\it i.e.} $W(x)=0$.  In
this case we simply get the $A_1$ geometry times the $x$-plane.  We
wrap $N$ branes around the ${\bf P}^1$.  In this case the normal
directions to the $P^1$ correspond to the cotangent bundle and the
trivial bundle ${\bf C}$ associated to $x$.  Let us call the Higgs
fields in these two direction respectively $\F_1(z)$ for the cotangent
direction and $\F_0(z)$ for the $x$-direction.  The topological theory on
the B-brane we obtain in this case is given by the action
$$
S = {1\over g_s}\int_{\P^1} \Tr\Bigl( \F_1 {\overline D_A} \F_0 \Bigr)
$$
where $\F_1(z)$ is a $U(N)$ adjoint valued $(1,0)$ form on ${\bf
P^1}$, $\F_0(z)$ is an adjoint valued scalar, $A(z)$ is a $U(N)$
holomorphic $(0,1)$ form connection on ${\bf P}^1$, and ${\overline
D_A}={\overline \partial}+[A,-]$.  Here $g_s$ denotes the string
coupling constant. If we turn on the Higgs fields thereby deforming the
$\P^1$ to a non-holomorphic curve ${\cal C}$, this action computes
the integral
$$
S={1\over g_s} \int_Y \Omega
$$
with $Y$ a 3-chain connecting $\P^1$ and ${\cal C}$.

If we turn on the superpotential $W$ by deforming the geometry, the
above action changes.  This has been studied in \kkl\ by studying the
Beltrami differential that deforms the complex structure in the
appropriate way, leading (including a minor generalization) to
\eqn\act{S_W = {1\over g_s}\int_{\P^1} \Tr\Bigl(
\F_1{\overline D_A}\F_0 +W(\F_0)\omega\Bigr).}
where $W(\F_0)$ corresponds to the superpotential and $\omega$ is a
$(1,1)$ form which can be taken to correspond to have unit volume on
${\bf P}^1$.

A consistency check for this action is to note that the equations of
motion following from the above action agree with the fact that the
classical solutions correspond to holomorphic curves.  In particular
integrating out $A$ gives
$$[\F_0,\F_1]=0,$$
so that $\F_0$ and $\F_1$ commute, {\it i.e.} we can assume they are
simultaneously diagonal.  Variation with respect to the eigenvalues of
$\F_1$ leads to
$${\overline \partial}\F_0=0,$$
which together with compactness of ${\bf P}^1$ implies that $\F_0$ is a
constant.  Variation with respect to $\F_0$ gives
$${\overline \partial}\F_1=W'(\F_0)\omega$$
which, together with the fact that the integral of ${\overline
\partial }\F_1$ over ${\bf P}^1$ has to be zero for non-singular
$\F_1$, leads to
$$
W'(\F_0)=0=\F_1.
$$
Thus the classical vacua indeed are localized at points $W'(\F_0)=0$,
which describe the positions of the $n$ $\P^1$'s..

In fact the action \act\ and the resulting quantum theory is rather
trivial.  In particular $\F_1$ also appears linearly and can be
integrated out exactly, leading to the constraint ${\overline \partial
\F_0}=0$, which leads to the statement that $\F_0$ is a constant
$N\times N$ matrix
$$
\F_0(z) = \F= cnst.
$$
Thus the full action just reduces to its last potential term,
which after integration of $\omega$ over ${\bf P}^1$ leads to the
matrix action
\eqn\matac{S_W(\F)={1\over g_s} \Tr\, W(\F).}
Thus we see that the partition function of the gauge system is
equivalent to a simple matrix model!

\subsec{Operator formalism}

We can give another derivation for the action \act\ starting directly
from the patching functions \patch\ that determine the blown-up
geometry. We split the $\P^1$ in two hemispheres connected by a long
cylinder.  We denote the fields on these two patches as $\F_0,\F_1$
respectively $\F'_0,\F'_1$. In the case $W=0$ we are simply dealing
with a gauged chiral CFT given by an adjoined valued $\beta$-$\gamma$
system of spin $(1,0)$. The partition function computes the number of
holomorphic blocks and this is one on the two-sphere. (Here we are ignoring
for a moment the factor coming from the volume of $U(N)$.)  In an
operator formalism this partition is simply given by pairing the left
and right vacuum
$$
Z = \langle 0 | 0 \rangle = 1.
$$
 Now we want to implement the deformation induced by
$W$. From the transition function \patch\ we see that the fields are
related in the following way (here we write the fields in coordinates on
the cylinder so that factors of $z$ and $z'$ are absorbed)
$$
\F_1' = \F_1 + W'(\F_0).
$$
Now there is an obvious operator $U$ that implements this
transformation on the Hilbert space. If we define
\eqn\smatrix{
U = \exp {1\over g_s} \oint {\rm Tr}\,W(\F_0(z)) dz,
}
(recall that the operator $\F_0(z)$ is an holomorphic field, so the
contour does not matter as long as it encircles the poles) then one
easily verifies that
$$
\F_1' = U \F_1 U\inv.
$$
Here one uses the fact that the fields $\F_0$ and $\F_1$ are
canonically conjugated
$$
\F_0(z) \F_1(w) \sim {g_s \over z-w}.
$$
Therefore in an Hamiltonian formalism the partition function should be
given by inserting the transformation $U$ between the left and right
vacua
$$
Z = \langle 0 | U | 0 \rangle.
$$
This is the familiar way to implement changes in complex structure in the
operator formalism of conformal field theory.

In equation \smatrix\ we have written the deformation of the action in
terms of the contour integration or Wilson line $\oint {\rm
Tr}\,W(\F_0)$. Alternatively, this can be written as a surface
integral
$$
U = \exp {1\over g_s} \int_{\P^1} {\rm Tr}\,W(\F_0)\, \omega,
$$
where  the volume form $\omega$ has been localized to a band
along the equator of the $\P^1$. But, as noted above, one can take
any 2-form, as long as it integrates to 1 over the sphere.

\subsec{Precise Conjecture}

We are now ready to state our conjecture in precise terms.  Let ${\cal
F}_{W,f}(g_s)$ denote the partition function of topological
B-model for the Calabi-Yau manifold given by
$$uv+y^2+W'(x)^2+f(x)=0,$$
where $W$ is a fixed polynomial of degree $n+1$ in $x$ and $f$ is a
polynomial of degree $n-1$ in $x$.  Let $S_i$ denote the integral of
the holomorphic 3-form over the $i$-th $S^3$ coming from the $i$-th
critical point of $W$.  The periods $S_i$ will vary as we vary the $n$
coefficients of $f$.  Inverting this map, given the variables $S_i$ we
can find the coefficients of the polynomial $f(x)$ compatible with these
periods.

On the gauge theory side we now consider the matrix model given by
the action
$$S_W(\F) ={1\over g_s}{\rm Tr}\, W(\F).$$
We expand this matrix model near the classical vacuum given by partitioning
$$
N = N_1 + \ldots + N_n,
$$
and by putting $N_i$ eigenvalues of $\F$ in the $i$-th critical point
of $W$. (Here we use both stable and unstable critical points. In
fact, since we work in the holomorphic context, this difference does
not really make sense.) Let ${\cal F}_{W,N_i}(g_s)$ denote the  free
energy of this matrix model expanded near this classical vacuum.  Then
the claim is
$${\cal F}_{W,f}(g_s)={\cal F}_{W,N_i}(g_s)$$
with the understanding that $N_i g_s=S_i$ and, as discussed above,
the periods $S_i$ fix the coefficients in the polynomial $f(x)$.

A few comments are in order: The matrix model integral is over a
holomorphic matrix $\F$, {\it i.e.} we have integrals of the form
$\int d\F$ and not like $\int d\F d{\overline \F}$. This is a general
issue in topological B-branes\foot{We thank E.~Witten for a useful
discussion on this point.}. The world-volume actions, such as the
holomorphic Chern-Simons action in six dimensions, is a
holomorphic function of the field variables. The non-perturbative
path-integral should therefore by defined by picking some appropriate
contour in the complex field configuration space. Indeed in our
two-dimensional example we ended up with a chiral CFT and the
path-integral definition of such a theory is rather subtle. One often ends up
defining it as a holomorphic square root of a non-chiral theory. For a
perturbative computation the situation is much easier. One simply
performs the Feynman diagrams as if one was dealing with a real field.
Similarly in this case we can effectively treat the matrix $\F$ as if
it is a real, that is Hermitean, matrix.

Secondly we should note that the matrix integral, in the particular
vacuum we end up, has a prefactor of
$$
1/{\rm Vol}(U(N_1)\times \cdots \times U(N_n)).
$$
This comes from the fact that for this vacuum $U(N_1)\times \cdots
\times U(N_n)$ denotes the unbroken gauge group and we have to mod out
by the corresponding volume of the constant gauge transformations.
This piece gives, as discussed in \ov\ the partition function of $c=1$
at self-dual radius.  In particular the genus $0$ answer will involve
$${\cal F}_0=\sum_i {1\over 2}S_i^2 {\rm log}S_i$$
In embedding in type IIB superstring this leads to the gaugino
superpotential $W_{eff}=\sum_i N_i{\partial {\cal F}_0}/\partial S_i
+\alpha S_i $.  (Note that within the type IIB context the parameters
$S_i$ and $N_i$ are independent.) This is a first check on our
conjecture.  We are now ready to test the above large $N$ conjecture
in more detail.

\newsec{Matrix models and special geometry}

We will first show how our conjecture can be proven in the planar limit
using standard manipulations in matrix model technology.  A useful
reference is for example \mmm. We will see how the special
geometry of Calabi-Yau's emerges naturally.

\subsec{Matrix integrals in the planar limit}

We consider the one-matrix integral over $N \times N$ Hermitean
matrices $\F$
$$
Z = {1\over {\rm Vol}(U(N))} \int d\F \cdot \exp\Bigl(
-{1\over g_s}{\rm Tr}\,W(\F) \Bigr)
$$
with $W(\F)$ a polynomial of degree $n+1$. By diagonalizing the matrix
$\F$ such integrals can be reduced to an integral over the eigenvalues
$\l_1,\ldots,\l_N$. In this way we obtain
$$
Z = \int \prod_i d\l_i \cdot \Delta(\l)^2 \cdot \exp
\Bigl({-{1\over g_s} \sum_i W(\l_i)}\Bigr),
$$
where the Vandermonde determinants
$$
\Delta(\l)=\prod_{i<j}(\l_i-\l_j) = \det\bigl(\l_i^{j-1}\bigr)
$$
appear from the Jacobian picked up by the diagonalization
process. After exponentiating this contribution the effective action
for the eigenvalues is given by
\eqn\action
{
S(\l) = {1\over g_s} \sum_i W(\l_j) - 2 \sum_{i<j} \log(\l_i-\l_j).
}
In this way we end up with a collection of $N$ variables $\l_1,\ldots,
\l_N$ in a potential $W(\l)$. These eigenvalues interact through the
second term that is a consequence of integrating out the off-diagonal
components. This term gives the famous Coulomb repulsion between the
eigenvalues.  Because of the effective Pauli exclusion principle
induced by the Vandermondes it makes the eigenvalues
 behave as fermions. From this action we see that the equation
of motion satisfied by the eigenvalues is
\eqn\eom{
{1\over g_s}W'(\l_i) - 2 \sum_{j \not=i} {1 \over \l_i -\l_j} = 0.
}

We will now take the limit $N \to \infty$ of this system while keeping
fixed the 't Hooft coupling
$$
\m = g_s N.
$$
In this standard large $N$ limit we will have a continuum of
eigenvalues and their density
$$
\rho(\l)= {1\over N} \sum_i \delta(\l-\l_i)
$$
becomes a continuous function on the real axis normalized to
$\int\rho(\l) d\l=1$. (In the following $\l$ will always denote a real
variable, in contrast with the variable $x$ that can be complex).  The
eigenvalues will fill a domain on the real axis. This domain might
consist of several disconnected components known as cuts. In the case
of more than one components one speaks of a multi-cut solution. In the
present case they are at most $n$ of these cuts. We will denote the
corresponding intervals in the complex $x$ plane as $A_i$,
$i=1,\ldots,n$.

\subsec{The spectral curve}

To further analyze the model it will be convenient (and standard
practice) to introduce the trace of the resolvent of the matrix $\F$
$$
\w(x) = {1\over N} {\rm Tr}\Bigl({1\over \Phi-x}\Bigr) =
{1\over N} \sum_i {1\over \l_i-x},\qquad x \in {\bf C},\ x \not= \l_i.
$$
This resolvent plays an crucial role in matrix model technology. It
also has an interesting physical interpretation. For
example, it can be thought of as a loop operator.

By multiplying the equation of motion \eom\ by the factor $1/(\l_i-x)$
and summing over $i$ one obtains the important relation (loop equation)
\mmm
\eqn\master{
 \w^2(x)- {1\over N} \w'(x) + {1\over \mu} \w(x)W'(x) + {1\over 4\mu^2}
f(x)=0 }
where the polynomial $f(x)$ is of degree $n-1$ and is given by
\eqn\fff{
f(x) = {4\mu \over N} \sum_i {W'(x) - W'(\l_i) \over x - \l_i}.
}
In some sense the function $f(x)$ determines through \master\ the
whole solution of the matrix integral. Since it is polynomial of
degree $n-1$ we only have to determine the $n$ unknown coefficients.

In the large $N$ limit the second term in \master\ can be ignored and
the differential equation for $\w(x)$ becomes an algebraic equation
\eqn\curve{
\w^2(x)+ {1\over \mu} \w(x)W'(x) + {1\over 4\mu^2}
f(x)=0 }
{}From this we see that the resolvent $\w(x)$ in general has a piece
that can have branch cuts. This singular part is captured by the
function $y(x)$ that we define here as
\eqn\yy{
\eqalign{
y(x) & = 2\mu \w(x)+ W'(x) \cr
     & = 2 g_s \sum_i{1\over \l_i-x} + W'(x). \cr
}}
In terms of the variables $(x,y)$ the relation \curve\ associated to
the matrix model now takes the form
\eqn\spectral{
y^2 - W'(x)^2 + f(x) = 0.
}
This has an interpretation as a hyperelliptic curve in the
$(x,y)$-plane.  We immediately recognize this curve as the curve
\cycurve\ that determined the periods and thereby the tree-level
topological string on the deformed Calabi-Yau geometry. In fact, it is
even more natural to identify directly the parametrization of the
curve in terms of $(x,\omega)$ in \curve\ with equation \omeg of
section 2. From this we see that we should identify $\w$ with the
field $z\F_1$.

Note that the function $y(x)$ naturally appears from the dynamics
of the eigenvalues. It is given by the variation of the action
\action\ with respect to a particular eigenvalue $\l$
\eqn\force{
y(\l) = g_s {\d S \over \d \l}.
}
In fact, from this we see that the quantity $y(x)$ has an elegant
physical interpretation. Being the derivative of the potential energy,
it equals the force on an eigenvalue if it tries to go away from its
stationary position and moves into the complex $x$-plane.

\subsec{Filling fractions and periods}

We now want to evaluate the matrix integral around a particular
stationary point where particular fractions of the eigenvalues
cluster around the different critical points. Around such a multi-cut
configuration it makes sense to make a perturbative expansion of the
matrix integral using large $N$ techniques. It is the contribution
from one of these saddle points that we are after.

Consider such a multi-cut solution. The filling fractions $N_i/N$,
{\it i.e.}\ the relative number of eigenvalues around each critical
point, are given by the integrals
$$
N_i/N = \int_{A_i} \rho(\l) d\l
$$
Here the $A_i$ are the cuts in the complex $x$-plane.

Now an important standard result is that the density of eigenvalues is
given by the jump of the resolvent $\w(x)$ across the cut
$$
\rho(\l) = {1\over 2\pi i} \Bigl(\w(\l+i0)-\w(\l-i0)\Bigr).
$$
Since only the singular piece contributes, we can
also write this relation as
\eqn\jump{
\rho(\l) = {\mu \over \pi i} \Bigl(y(\l+i0)-y(\l-i0)\Bigr).
}
Therefore we can compute the fraction of eigenvalues in a specific
cut by doing a contour integral around the cut. In this way we find
that the quantities $S_i=g_sN_i$ are given by the period integrals
around the cut, that is the periods on the Riemann surface
\eqn\Aperiod{
S_i= {1\over 2\pi i} \oint_{A_i} y(x) dx.
}
Here we make contact with the period integrals \periods\ as obtained
in the topological B-model computation.  This is the first half of the
derivation of the genus zero part of our conjecture.

In order to complete the derivation, we now need to compute the change
in the free energy $\cF_0(S_i)$ if we vary the filling fractions $S_i$
by adding an eigenvalue to the cuts
$$
\Delta S_i = {1\over g_s} \Delta N_i.
$$
This change in the action is given by the work
done by the force $F(x)$ acting on an eigenvalue if we move this eigenvalue
from one branch to infinity. We have seen in \force\ that this force
is given by
$$
F(x) = {1\over g_s} y(x).
$$
So the variation of the free energy is computed in terms of the action
$\int F(x) dx$, that is by integrating the one-form $y(x) dx$ along
one of the $B$-cycles to the cut-off point $x=\Lambda$. We therefore
immediately find the special geometry relation
\eqn\Bperiod{
{\partial \cF_0 \over \partial S_i} = \int_{B_i} y(x) dx
}
expressing the $B$-periods in terms of the $A$-periods through the
free energy $\cF_0(S)$. Together \Aperiod\ and \Bperiod\ give the
precise match with the Calabi-Yau geometry. This concludes the
derivation of the planar version of our conjecture relating the matrix
model to the topological B-model on the deformed Calabi-Yau.

As we have mentioned, for a given potential $W(x)$ the function $f(x)$
that deforms the singular Calabi-Yau and determines the solution of
the matrix model can be expressed in terms of the periods $S_i$ and
vice versa. In fact, using definition \fff\ we can give a useful
relation valid in terms of the prepotential $\cF_0$. If one
parametrizes the potential as
$$
W(x) = \sum_{k=0}^{n+1} u_k x^k,
$$
and the deformation as
$$
f(x) = \sum_{k=0}^{n-1} 4 \mu b_k x^k,
$$
then by plugging definition \fff\ in the matrix integral we obtain
$$
b_k = (k+2)u_{k+2} + \sum_{j=k+2}^{n+1} j u_j {\d \cF_0(u_i,S_i) 
\over \d u_{j-k-2}}.
$$
Note that if we introduce Virasoro operators $L_k = \sum_j j u_j \d/\d
u_{j+k}$ this equation can be written as
\ref\vir{M. Bertola, B. Eynard, J. Harnad, ``Partition functions for Matrix
Models and Isomonodromic Tau functions,'' arxiv:nlin.SI/0204054.}
$$
b_k =(k+2)u_{k+2} + L_{-k-2}\cF_0,\qquad k=0,\cdots,n-1.
$$

\subsec{Higher genus}

Can one extend the derivation of our conjecture to higher genera? One
possibility might be to use the loop equations that are derived by
taking the expectation value of expression \master\ for the resolvent
$\w(x)$
$$
\langle \w^2(x) \rangle + {1\over \mu} \langle \w(x)\rangle \,
W'(x) + {1\over 4\mu^2}f(x)=0,
$$
where $f(x)$ is now given by the expectation value of expression \fff.
These loop equations have been studied for multi-cut solutions, for
example in \loop, and one can try to solve them order by order in the
string coupling constant $g_s$. In principle these equations give
recursion relations that relate the higher genus amplitudes in terms
of the tree-level free energy. It would be very interesting to see if
these equations are directly related to the equations of motion for
Kodaira-Spencer string field theory \bcov. This is not completely
unlikely since collective field theories for the eigenvalue densities
are known to have a similar form \dj, and morally the connection
between topological strings and matrix models should go along these
lines.

It might be interesting to also briefly discuss the case of the pure
conifold, here given by the quadratic superpotential $W(x)=x^2.$ The
large $N$ dual is from our point of view just the Gaussian matrix
model
$$
Z = {1\over {\rm Vol}(U(N))} \int d\F \cdot e^{-{1\over g_s}\Tr\,\F^2}.
$$
Since the integral is trivial, the only contribution comes from
the normalization factor
$$
 {1\over {\rm Vol}(U(N))} \sim \prod_{k=1}^{N-1} k!
$$
which has been shown to reproduce the all genus answer for the B-model
on the conifold in the $1/N$ expansion \ov.  In this case the spectral
curve is given by
$$
y^2-x^2+\m=0
$$
and the eigenvalue density
$$
\rho(\l)= \sqrt{\l^2 - \m}
$$
is Wigner's famous semi-circle distribution. In the eigenvalue basis
the all genus answer is alternatively obtained by the method of
orthogonal polynomials which indeed gives
\ref\mor{A.~Morozov, ``Matrix Models as Integrable Systems,''
arxiv:hep-th/9502091.}
$$ 
\int \prod_i d\l_i \cdot \Delta(\l)^2 \cdot e^{
-{1\over g_s} \sum_i \l_i^2} \sim \prod_{k=1}^{N-1} k!.
$$

\subsec{Domain walls and eigenvalue tunneling}

There is an interesting relation that directly connects the
D-branes in the type II string theory and the behaviour of the
eigenvalues in the matrix models. In the ``old days'' it was pointed
out by Shenker \shenker\ that the characteristic non-perturbative
effect observed in matrix models was the tunneling of eigenvalues and
this was an effect of strength $1/g_s$. This remark in some sense
anticipated the importance of D-branes. Here we can connect
the two effects.

In the type IIB theory on the resolved geometry we can consider
D5-branes wrapped around an $S^3$ that interpolates between two
$S^2$. Such an object manifest itself as a domain wall in the four
uncompactified spacetime dimensions. After the geometric transition
such a D5-brane will connect two three-cycles. It will describe a
process where one unit of RR flux is transported. That is, the RR flux
in one $S^3$ is decreased by one unit and the flux in another $S^3$
is increased by one. Since the
space-time superpotential is given by
\eqn\superpotential{
W_{eff} = \sum_i \Bigl(N_i {\d \cF_0 \over \d S_i} + \a S_i \Bigr),
}
the tension of a domain wall transferring flux from the $i$-th to the
$j$-th cycle is given by
$$
T =   {\d \cF_0 \over \d S_i} -  {\d \cF_0 \over \d S_j}=
\int_{B_{ij}} y(x) dx.
$$
We now recognize this as the instanton action in the matrix model of
an eigenvalue tunneling from the cut $A_i$ to the cut $A_j$ along the
path $B_{ij}$.

\newsec{Double scaling limits and other further generalizations}

Given that we have found a natural stringy interpretation
of ordinary matrix model, one could ask what is the meaning
of the double scaling limit in the context of the old matrix model
\mmm.
Following that limit on the gravity side for the single matrix model
leads to the Calabi-Yau geometry
$$H=uv+y^2+x^{2m+1}+{\rm deformations}=0$$
as corresponding to the $(2,2m+1)$ bosonic minimal model coupled to
two-dimensional gravity.  The deformations correspond to the $m$
observables of the $(2,2m+1)$ model. In fact for generic deformations
of this geometry there are $m$ A-cycles and $m$ B-cycles and we thus
can choose $m$ independent parameters to parametrize these deformations.
The infinitesimal deformations which map this geometry to the
deformations of the $(2,2m+1)$ models are of the form
\eqn\lge{uv+y^2+x\prod_{i=1}^m (x-\epsilon_i)^2=0,}
where $\epsilon_i$ are related to the deformation of the $(2,2m+1)$
model with primary fields.  For example, the case of pure 2d gravity,
{\it i.e.} the $(2,3)$ model has only one observable and in that case we
have $\epsilon =\mu^{1/2}$.  In the usual matrix model this is
obtained by considering a quartic superpotential $W(x)$.  Let us take
that to be an even function of $x$.  Then there are three critical
points, including one at $x=0$.  Upon deformation by $f(x)$ the
critical points can split.  If we put all the $N$ eigenvalues at the
well corresponding to $x=0$ then only the $x=0$ double point splits
and the other two do not split.  The double scaling limit corresponds
to taking the limit where one of the double points reaches one pair of
doubles and the other reaches the other pair. Taking the limit where
the geometry is localized near one set of triple zeros we obtain the
$(2,3)$ local geometry given in \lge.  Embedding this kind of
theory in type IIB strings gives exactly the kind of $N\rightarrow
\infty$ dualities proposed recently in  \amv\
which relate certain limits of gauge systems with type IIB strings on
Calabi-Yau geometries without fluxes.

Note that this map to topological B-model is in perfect accord with
the fact that the bosonic strings have a hidden ${\cal N}=2$
superconformal symmetry \berwa, which in this case we can identify
with the superconformal theory on the above Calabi-Yau threefold.  As
a check note that, if we only turn on the cosmological constant, then
the genus $g$ answer for topological B-model scales as the holomorphic
threeform $\Omega$ to the power of $2-2g$ \bcov.  For the $(2,3)$
model since we have
$$\Omega ={dudvdydx\over dH}\sim {\mu}^{5/4}.$$
So we learn that the free energy of the B-model is given by
an expansion of the form
$${\cal F}=\sum_{g \geq 0} c_g g_s^{2g-2} {\mu}^{{5\over 4}(2-2g)}$$
for some coefficients $c_g$, in perfect accord with the expected answer.

As a further generalization one can consider the topological lift of
the models considered in \cfikv .  These involve many $U(N)$ gauge
fields with bifundamental matter.  In particular for the case of a
linear chain of $p-1$ gauge fields one obtains the dual geometry
(before blow up)
$$uv+\prod_{i=1}^p (y-W_i'(x))=0.$$
where the $W_i(x)$ are related to the superpotentials of the
corresponding gauge fields.  Again a double scaling limit exists which
lead to geometries of the form
$$uv+y^p+x^q+{\rm deformations}=0$$
This would correspond to the $(p,q)$ bosonic minimal model coupled to
gravity.  There are ${1\over 2}(p-1)(q-1)$ A-cycles for deformations
in the above equation, which correspond to the observables of the
$(p,q)$ minimal model. As discussed in \ckv\ in this case the
reduction of holomorphic 3-form leads to $p-1$ 1-forms $\eta_i$ which
naturally get identified with the $p-1$ eigenvalue densities in a
$(p-1)$ matrix model.  It would be interesting to study these model in
more detail and in particular the corresponding multi-matrix model
duals.

\subsec{Final remarks}

There are many indications that topological strings on a general
target space might be described by some kind of integrable
systems. This was originally shown for the $c\leq 1$ topological
string theories \dijk. More recently in the context of A-model
topological strings on non-compact Calabi-Yau this integrability was
shown by dualizing to certain observables of Chern-Simons gauge theory
on three-manifolds \refs{\amv ,\dfg}.  For A-models on compact target
spaces evidence has been accumulating in the mathematical literature
on Gromov-Witten invariants (see for example \refs{
\givental,\op}). If we manage to formulate a matrix model dual of the
topological string this integrability is in some sense manifest. Both
the finite $N$ and the large $N$ matrix models are well-known to give
tau-functions of the KP and Toda hierarchies
\refs{\mmm,\mor}. This integrability of matrix models was the
underlying reason that non-critical strings with $c \leq 1$ were
exactly solvable. Our results indicate that for a large class of
non-compact Calabi-Yau manifolds this integrability is present.

Furthermore one should also like to be able to include gravitational
descendents within the topological string model. In terms of the
B-model we expect that these are given by non-normalizable
deformations of the complex structure.

A related issue is the description of the $c=1$ string, which is
equivalent to topological strings on the conifold geometry \ghv . In a
forthcoming paper \cone\ we will describe how tachyon scattering
processes are indeed reproduced in the conifold string theory and can
be described by a large $N$ dual gauge system, making contact with
\wiegmann\ and the recent work \kazakov.

\bigskip
\centerline{\bf Acknowledgements}

We would like to thank F.~Cachazo, V.~Kazakov, G.~Moore and H.~Ooguri
for valuable discussions. R.D. would like to thank the Harvard Physics
Department and the Institute for Advanced Study, Princeton for kind
hospitality during part of this work. The research of R.D. is partly
supported by FOM and the CMPA grant of the University of Amsterdam,
C.V. is partly supported by NSF grants PHY-9802709 and DMS-0074329.

\listrefs

\bye